\documentstyle[11pt,amssymb]{article}

\makeatletter
\@addtoreset{equation}{section}
\makeatother

\def\dalemb#1#2{{\vbox{\hrule height .#2pt
        \hbox{\vrule width.#2pt height#1pt \kern#1pt
                \vrule width.#2pt}
        \hrule height.#2pt}}}

\def\cM{{\cal M}}

\def\ep{\epsilon}
\def\td{\tilde}

\def\half{{\textstyle{1\over2}}}

\def\qu{{\textstyle{1\over 4}}}
\def\tq{{\textstyle{1\over 6}}}

\def\tw{{\textstyle{1\over 12}}}
\let\a=\alpha  \let\g=\gamma \let\d=\delta \let\e=\epsilon
  \let\q=\theta  \let\k=\kappa
\let\l=\lambda \let\m=\mu \let\n=\nu  
 \let\t=\tau  \let\f=\phi  
\let\w=\omega    
    
    \let\G=\Gamma
\let\la=\label  
  
\def\nn{\nonumber} \def\bd{\begin{document}} \def\ed{\end{document}}
\def\ds{\documentstyle} \let\fr=\frac \let\bl=\bigl \let\br=\bigr
\let\Br=\Bigr \let\Bl=\Bigl
\let\bm=\bibitem
\let\na=\nabla
\let\pa=\partial \let\ov=\overline
\newcommand{\be}{\begin{equation}}
\newcommand{\ee}{\end{equation}}
\def\ba{\begin{array}}
\def\ea{\end{array}}
\def\ft#1#2{{\textstyle{{\scriptstyle #1}\over {\scriptstyle #2}}}}
\def\fft#1#2{{#1 \over #2}}
\def\del{\partial}
\def\sst#1{{\scriptscriptstyle #1}}
 \def\oneone{\rlap 1\mkern4mu{\rm l}}
\def\ie{{\it i.e.\ }}
\def\via{{\it via}}
\def\semi{{\ltimes}}

\def\sp{\; \; \;}

\def\bol{ \left | B \right \rangle}
\def\bo1{ \left | B' \right \rangle}

\newcommand{\hsp}{\hspace{0.5cm}}

\newcommand{\ho}[1]{$\, ^{#1}$}
\newcommand{\hoch}[1]{$\, ^{#1}$}
\newcommand{\bea}{\begin{eqnarray}}
\newcommand{\eea}{\end{eqnarray}}
\newcommand{\ra}{\rightarrow}
\newcommand{\lra}{\longrightarrow}
\newcommand{\Lra}{\Leftrightarrow}
\newcommand{\ap}{\alpha^\prime}
\newcommand{\bp}{\tilde \beta^\prime}
\newcommand{\tr}{{\rm tr} }
\newcommand{\Tr}{{\rm Tr} }

\newcommand{\co}{{\cal O} }
\newcommand{\ca}{{\cal A} }
\newcommand{\cb}{{\cal B} }

\newcommand{\spin}{{\it Spinoza Institute, University of Utrecht,\\
Postbus 80.195, 3508 TD Utrecht, The Netherlands}\\
{\tt email:taylor@phys.uu.nl}}
\newcommand{\ams}{{\it Institute for Theoretical Physics,
University of Amsterdam, \\
Valckenierstraat 65, 1018XE Amsterdam, The Netherlands} \\
{\tt email:skenderi@science.uva.nl}}
\newcommand{\dan}{{\it Department of Mathematics and Center for
Theoretical Physics, \\
Massachusetts Institute of Technology, Cambridge MA 02139, USA} \\
{\tt email:dzf@math.mit.edu}}
\newcommand{\auth}{\large\bf{Daniel Z. Freedman\hoch{\ast},
Kostas Skenderis\hoch{\star} and Marika Taylor\hoch{\dagger}}}

\begin{document}

\begin{flushright}
\hfill{\bf hep-th/0306046}\\
\hfill{SPIN-2003/17} \\
\hfill{ITF-2003/26} \\
\hfill{ITFA-2003-26 \\
MIT-CPT/3382}
\end{flushright}

\vspace{15pt}

\begin{center}

{\Large \bf Worldvolume supersymmetries for branes in plane waves}

\vspace{20pt}

\auth
\vspace{15pt}

{\hoch\ast \dan}

\vspace{8pt}

{\hoch\star \ams}

\vspace{8pt}

{\hoch\dagger \spin}

\vspace{15pt}

\underline{ABSTRACT}
\end{center}
We study the worldvolume supersymmetries of M2 branes in the maximally
supersymmetric plane wave background of M theory. For certain embeddings
the standard probe analysis indicates that the worldvolume theory
has less than 16 supersymmetries. We show that at the quadratic level
the worldvolume theory admits additional linearly realized supersymmetries,
and that the spectra of the branes are organized into multiplets
of these symmetries. We find however that these supersymmetries
are not respected by worldvolume interactions. Our analysis was motivated by
recent work showing that D-branes in the maximally supersymmetric
plane wave background of IIB string theory admit supersymmetries
beyond those of the probe analysis. The construction
of the additional supercharges in this case was specific to 
a string worldsheet that is a strip and the present results suggest
that string interactions do not preserve these symmetries. 

\noindent

\pagebreak
\setcounter{page}{1}

\tableofcontents
\addtocontents{toc}{\protect\setcounter{tocdepth}{2}}


\section{Introduction}

One of the most elementary questions that one can ask about a
supersymmetric theory is ``what are the supersymmetric states of
the theory''. In superstring theory a class of supersymmetric
states is represented by D-branes. Recent work on branes in the
maximally supersymmetric plane wave background of IIB string theory
shows that this issue is more subtle than the corresponding
analysis in flat spacetimes. A tree-level open string analysis 
of boundary conditions and spectra in
\cite{ST2,ST3} revealed that certain branes have more
supersymmetries than the probe analysis gives \cite{ST1}. This
implies that either the standard probe analysis needs to be
amended or that the extra supersymmetries are not respected by
string interactions. One of the aims of the present work is to
settle this issue.

Recall that in string perturbation theory D-branes specify
boundary conditions for open strings. In the Green-Schwarz
formalism the spacetime supersymmetries preserved by the D-brane
manifest themselves as global symmetries of the worldsheet action.
Some of the global symmetries of the open worldsheet directly
descend from corresponding symmetries of the closed string. These
symmetries are exactly the ones found by the probe analysis. It
was found in \cite{ST2,ST3} however that in certain cases the
worldsheet action admits additional supersymmetries and that the
spectrum of the theory is organized with respect to the extra
supersymmetries as well.

The branes in the IIB plane wave can be divided into two different
classes: the $D_-$ and $D_+$ branes. This classification
originates from the specific form of the (worldsheet) fermionic boundary
conditions, but one can also understand it from the features of the
spectrum. In $D_-$ branes, the mass parameter $\mu$ of the plane
wave lifts some of the degeneracy of states present in the flat
space limit. In particular, the lowest lying states which in the
flat space limit form a $d=10$ vector multiplet (i.e. a multiplet
with 8+8 degrees of freedom with the same lightcone energy
$P^-$) now splits as 1+4+6+4+1 (see Table 1
of \cite{ST3}). On the other hand, the lowest lying 
states for $D_+$ branes are as degenerate
as in flat space (see Table 3 of \cite{ST3}).

The branes under investigation are located at a constant
transverse position and wrap specific directions. According to the
probe analysis, $D_+$ branes always break all kinematical
supersymmetries\footnote{We call kinematical supersymmetries the
supercharges that square to the lightcone momentum, and dynamical
the ones that square to the lightcone Hamiltonian (plus other
charges).}. $D_-$ branes preserve 8 kinematical supercharges
along with an additional 8 dynamical supercharges only
when the brane is located at the origin \cite{ST1}. In the string theory
analysis \cite{ST2}, however, one finds 8 alternative supercharges 
for the $D_{+}$ branes and for the $D_{-}$ branes located away
from the origin. A clue for the existence of the extra supersymmetries
was that the spectrum of the brane exhibits more
symmetries than the probe analysis suggests. In particular, the
spectrum of $D_-$ branes at and away from the origin is identical
up to an overall additive (positive) constant in the lightcone
energy \cite{ST3}. 

Notice that the existence of extra {\it local} Noether currents
corresponding to the new supersymmetries is not an automatic
consequence of the degeneracy of the spectrum.
Indeed, the string states of the $D_+$ branes also
appear to be organized in multiplets of ``dynamical supercharges''. 
However, these charges are associated with {\it non-local}
currents \cite{ST3}.

The extra 8 supersymmetries for $D_+$ branes are completely new
symmetries, unrelated to the closed string symmetries. They
satisfy the standard lightcone superalgebra, i.e. they square to
the lightcone momentum. The extra 8 supersymmetries for $D_-$
branes are a combination of the corresponding closed string
supersymmetries and new transformation rules. The new supercharges
when evaluated on-shell are identical to the corresponding
supercharges of the brane at the origin \cite{ST3}. It follows
that the corresponding superalgebras are also identical (up to
certain c-number shifts). In particular, the new supercharges
square to the lightcone Hamiltonian plus rotational charges 
plus a c-number that has the interpretation of the energy of
an open string in a harmonic oscillator potential with ends at the
constant position of the brane. This c-number contribution 
is also the on-shell value of an additional worldsheet charge 
(see (4.31)-(4.32) of \cite{ST2}).

The construction of the extra supercharges crucially  used the
fact that the gauge fixed worldsheet action is quadratic in the
fields and that the worldsheet is a strip. The former is a special
property of strings propagating in the IIB plane wave background.
The latter indicates that the extension of the extra symmetries to
higher genus surfaces is not immediate, and that string
interactions may invalidate them.

Consistency requires that the string theory and probe
analysis yield the same results. Recall that the worldvolume
theory of D-branes captures the low-lying open string excitations
and their (low-energy) interactions. In particular, the spectrum
of small fluctuations around the D-brane embedding should coincide
with the zero slope limit of the open string spectrum. As
mentioned, the open string spectrum is more supersymmetric than
the probe analysis implies. It follows that the quadratic part of
the D-brane action should exhibit additional supersymmetries. If
string interactions respect the extra symmetries then the
worldvolume interactions should also respect them. Conversely, if
we show that the worldvolume interactions do not preserve the
extra symmetries then this shows that string interactions do not
respect the extra symmetries.

Recall that the brane worldvolume theories are by construction
invariant under target space supersymmetry and they also possess
a local kappa symmetry invariance. Upon gauge fixing the kappa symmetry,
the target space supersymmetry turns into worldvolume supersymmetry.
The plane wave backgrounds we discuss in this paper admit 32 supercharges.
This means that the worldvolume theory is by construction
invariant under 32 fermionic symmetries. However, at most 16 of them
are linearly realized, i.e. they are of the schematic form
(the exact expressions are given in the main text)
\be
\d X^A \sim  \bar{\theta} \G^A \e + \cdots , \qquad
\delta \theta \sim \G^{\mu A} \pa_\mu X^A \e + \mu \G^A X^A \e + \cdots
\ee
where $\e$ is the supersymmetry parameter and the dots indicate 
additional terms which are at least quadratic in the fields.
The remaining transformations have an inhomogeneous term
\be
\delta \theta = \beta + \cdots
\ee
where $\beta$ is field independent, and the associated $\theta$ can
be identified
with the Goldstone fermion associated with the breaking of supersymmetry.

The probe analysis by construction counts the number of linearly realized
supersymmetries that arise from a combination of target space supersymmetries
with kappa symmetry. This does not exclude however the possibility that
there are extra non-generic symmetries when the brane is in particular
backgrounds. The string theory analysis in \cite{ST2,ST3} can be
viewed as an example of such phenomenon: because of special properties
of the background (i.e. the worldsheet action is quadratic in fields)
the worldsheet theory exhibits more symmetries than in generic situations.

The extra linearly realized supersymmetries can be
completely new symmetries, unrelated to the supersymmetries associated with
target space supersymmetries. Alternatively, if the brane in
special backgrounds exhibits a new gauge invariance that allows one to
gauge away the Goldstone fermion then the corresponding symmetry
would be linearly realized. Both mechanisms are suggested by the
string theory computation in \cite{ST2}: the former is the analogue
of the new kinematical supersymmetries in $D_+$ branes and the
latter is the analogue of the restoration of dynamical supersymmetries
for $D_-$ branes using worldsheet symmetries. We will see that
both mechanisms are realized, albeit only at the quadratic
approximation of the worldvolume theories.

The purpose of this paper is to analyze the issue of worldvolume
supersymmetries in detail. Instead of working with the worldvolume
theory of IIB D-branes, however, we will analyze the same issues
for M2 branes for which the worldvolume theory is much simpler (since there 
are no gauge fields). The probe analysis for this case has been worked
out in \cite{KY} (see also \cite{MR}). The results for
supersymmetric M2 embeddings are directly analogous to the results in
\cite{ST1}. Recall that in the maximally supersymmetric plane wave
of M-theory the transverse to the lightcone coordinates split as
3+6. The M2 branes that wrap the lightcone directions and one of
the 3 coordinates preserve 16 supercharges when located at the
origin of transverse space but only 8 when located away from it.
These are the analogues of $D_-$ branes and we will refer to them
as $M_-2$ branes. M2 branes that wrap the lightcone coordinates
and one of the 6 coordinates preserve no supersymmetry and are the
analogue of $D_+$ branes. They will be referred to as $M_+2$
branes. We will see here that their fluctuation spectra are
similar to those of the corresponding $D_-$ and $D_+$
branes.

Of course in this case there is no computation corresponding to the string 
theory analysis in \cite{ST2,ST3} so strictly speaking there is 
no a priori reason for expecting the worldvolume theory to 
exhibit extra supersymmetries. 
Since the probe computations for M and D branes are
similar, we expect extra supersymmetries at the quadratic level of
fluctuations of the M2 probe action, and our calculations confirm
this. Conversely our results in \S\ref{m-sec} and \S\ref{m+sec} 
indicate that the extra
supersymmetries fail at the interacting level, and we expect the
same to be true for D-branes.

This paper is organized as follows. We review the properties
of the plane wave background in \S\ref{pw} and discuss the
computation of the gauge fixed supermembrane action in \S\ref{m2sec}. 
In \S\ref{m-sec} we consider the spectrum and worldvolume
supersymmetries of $M_-2$ branes whilst section \S\ref{m+sec}
addresses the same issues for $M_{+}2$ branes. In the discussion
section \S\ref{dis} we review our results and comment on
the implications for D-branes in type IIB plane waves
and for the stability of the branes. 

\section{The plane wave background} \label{pw}

The maximally supersymmetric plane wave background
of eleven-dimensional supergravity is \cite{K}
\bea
ds^2 &=& - 2 dx^+ dx^- + \sum_{A =1}^{9} (dx^A)^2 - 
\frac{\mu^2}{36} 
\left(4 \sum_{i=1}^3 (x^i)^2 + \sum_{a=4}^9  (x^a)^2 \right) (dx^+)^2;
\nn \\
F_{+123} &=& - \mu \qquad \longrightarrow \qquad C_{ijk} = - \mu x^+ \ep_{ijk},
\eea
where a specific gauge choice for the 3-form gauge gauge field $C$ is 
made.

For coset spaces such as the plane wave
exact expressions for the supervielbein were computed in \cite{WPPS}
(see also \cite{WPP})
\bea \label{superv}
E &=& D \q + \sum_{n=1}^{16} \frac{1}{(2n+1)!} {\cM}^{n} D \q;  \\
E^r &=& e^r + \bar{\q} \G^r D \q + 2 \sum_{n=1}^{15} \frac{1}{(2n+2)!}
\bar{\q} \G^{r} \cM^{n} D \q. \nn
\eea
Here $(r,\bar{a})$ are tangent space vector and spinor
indices, respectively, and $(m,\a)$ are the corresponding curved indices.
In these expressions{\footnote 
{Note that we will use here the usual WZ gauge for the 
plane wave superspace, namely $\q^{\bar{a}} = 
\delta^{\bar{a}}_{\sp \alpha} \q^{\alpha}$. An alternative choice
would be the Killing spinor adapted gauge \cite{KC} 
$\q^{\bar{a}} = K^{\bar{a}}_{\sp \a} \q^{\a}$ where
the Killing spinors of the curved target space are
$\ep^{\bar{a}} = K^{\bar{a}}_{\sp \a} \ep_{0}^{\a}$ with constant 
$\ep_{0}^{\a}$.
In contrast to the $AdS$ backgrounds considered in \cite{KC} this
choice does not appear to lead to substantial simplifications in
the supervielbeins or the worldvolume action.}}
\bea \label{sv}
D \q &=& d \q + e^r T_{r}{}^{stuv} \q F_{stuv} + \qu \w^{rs} \G_{rs} \q; \\
T_{r}{}^{stuv} &=& \frac{1}{2! 3! 4!} 
(\G_{r}{}^{stuv} - 8 \d_r^{[s} \G^{tuv]}); \nn \\
\cM &=& 2 (T_r{}^{stuv} \q) F_{stuv} (\bar{\q} \G^r)
- \frac{1}{288} (\G_{rs} \q) (\bar{\q} [\G^{rstuvw} F_{tuvw} + 24 \G_{tu}
F^{rstu}] ). \nn
\eea
The bosonic vielbein and the spin connection are 
\be
e^{\sp -}_{-}  = e^{\sp +}_{+ } = 1, \hsp
e^{\sp A}_{B} = \d^{\sp A}_{B}, \hsp
e^{\sp -}_{+} = -\half G_{++}, \hsp \w_{+}{}^{-A} = -\half \pa_{A} G_{++}.
\ee
($G_{mn}$ denotes the spacetime metric). The Killing spinors are
\be \la{kil}
\ep = \left(1 
+ \frac{\mu}{12} (x^{a} \G^{a}- 2 x^i \G^i) \G^{+123} \right) 
\exp\left(\frac{\mu}{12} x^+ \G^{+-123}\right) 
\exp\left(-\frac{\mu}{6} x^+ \G^{123} \right) 
\ep_0
\ee 
where $\ep_0$ is a constant spinor.

\section{Supermembrane action} \label{m2sec}

The supermembrane action \cite{BST} is
\be \label{m2a}
S = - \int d^3 \xi \sqrt{- {\rm{det}} g_{\mu \nu} } + \int B,
\ee
where the induced worldvolume supermetric is
$g_{\mu \nu} = \Pi^{r}_{\mu} \Pi^{s}_{\nu} \eta_{rs}$ and
$\Pi^{r}_{\mu} = \pa_{\mu} Z^{M} E^{r}_{M}$. Here $Z^M=(X^m, \q^{\a})$
are the coordinates of the target superspace and $\xi^{\mu}$
are the worldvolume coordinates.
The explicit expression for $B$ in the coset background is
\cite{WPPS}
\be
B = \frac{1}{6} e^{r} \wedge e^s \wedge e^t C_{rst}
- \int^{1}_{0} dt \bar{\q} \G_{rs} E(t) \wedge E^{r}(t) \wedge E^s(t),
\ee
where $t$ is auxiliary and $E(t)$, $E^{r}(t)$ are obtained from
the supervielbeins by the shift $\q \rightarrow t \q$.

This action is invariant under the kappa symmetry transformations \cite{BST}
\be \la{kap}
\delta Z^M E_M^r = 0, \hsp
\delta Z^M E^a_M = [(1 - \Gamma) \kappa]^a,
\ee
where
\be
\Gamma = {1 \over 6} \frac{\ep^{\mu \nu \rho}}{\sqrt{-g}} \Pi^{r}_{\mu}
\Pi^s_{\nu} \Pi^t_{\rho} \G_{rst}
\ee
which satisfies $\G^2 =1$ and ${\rm{Tr}}({\G}) = 0$. The action is
also invariant under superspace diffeomorphisms
$\delta Z^{M} = - K^{M} (Z)$ which act as
\bea \la{sup}
\delta E_{M}^{\sp A} &=& K^{N} \pa_{N} E_{M}^{\sp A} +
\pa_{M} K^{N} E_{N}^{\sp A}; \\
\delta B_{MNP} &=& K^{Q} \pa_{Q} B_{MNP} + 3 \pa_{[M} K^Q B_{|Q| NP]}, \nn
\eea
where $K^{M}(Z)$ is a Killing supervector,
along with worldvolume diffeomorphisms 
\be \la{wd}
\delta Z^{M} = \eta^{\mu} \pa_{\mu} Z^{M}; \hsp
\delta g_{\mu \nu} = \eta^{\rho} \pa_{\rho} g_{\mu \nu}
+ 2 \pa_{(\mu} \eta^{\rho} \g_{\nu) \rho}.
\ee
The action also admits other symmetries, such as tensor gauge
transformations, but these will not play a role here.
To leading order the kappa symmetry and supersymmetry transformations
of the supermembrane action are
\bea \la{supkap}
\delta_{\kappa} \q = (1 - \Gamma) \kappa, \hsp
\delta_{\kappa} X^{m} = - \bar{\q} \G^{m} \d_{\k} \q; \\
\delta_{\ep} \q = \ep, \hsp
\delta_{\ep} X^{m} = - \bar{\ep} \G^{m} \q,
\eea
where $\ep$ are the Killing spinors of the target space (\ref{kil}).
Implicit in (\ref{kap}) and (\ref{sup}) are corrections to these
expressions which are higher
order in $\theta$ and can be neglected in what follows.

\subsection{Embeddings}

Membrane embeddings are given by solutions of the bosonic field equations
of (\ref{m2a}),
\be
{1 \over \sqrt{-\g}} \pa_\m (\sqrt{-\g} \g^{\m \n} \pa_\n X^m) 
+ \g^{\m \n} \pa_\m X^n \pa_\n X^p \G^m_{np} = 
{1 \over 3!} \e^{\l \m \n}F^m{}_{\l \m \n}
\ee
where $\g_{\mu \nu}=\pa_\m X^m \pa_\n X^n G_{mn}$ is the induced
worldvolume metric and
$\G^m_{np}$ is the Christoffel symbol of the plane wave metric.

The solutions of interest here have been discussed in  \cite{KY},
so our discussion will be brief. These solutions are
\bea 
M_-2:&& \quad 
 X^{\mu} = \xi^\mu,\ \m=\{+,-,1\}, \qquad X^{A'}=x_0^{A'}, 
\ A'=\{i',a\}  \label{embed} \\
M_+2:&& \quad
X^{\mu} = \xi^\mu,\ \m=\{+,-,4\}, \qquad X^{A'}=x_0^{A'},  
\ A'=\{i,a'\}  \nn
\eea
where in each case $A'$  runs over directions transverse to the brane,
$i'=2,3$ and $a'=5,..,9$. 

The supersymmetries of branes at the origin {\it vs.} branes
displaced along the parabolic ``potential'' in the transverse
directions are one of the main concerns of this paper. Although the
plane wave space-time is homogeneous, rigid translations in
transverse directions are not isometries. So branes at $x_0^{A'}=0$
and $x_0^{A'} \ne 0$ are not related by symmetry and are
physically distinct.

The condition for unbroken supersymmetry is \cite{BDPS}
\be
0=\delta \q = \ep (X) + (1 -\g^*) \kappa (X)
\ee
where $\g^*$ is $\G$ evaluated at the embedding and $\e$ is the Killing
spinor of the background evaluated at the embedding. Clearly one
can choose $\kappa (X)$ to cancel the effect of the $(1- \g^*)
\ep (X)$ projection of the Killing spinor, so the condition
reduces to
\be \label{sucon}
\g^* \e = -\e
\ee

For $M_- 2$ branes, $\g^* = \G_{+-1}$, and for $M_+ 2$ branes
$\g^* = \G_{+-4}$. Using the decomposition of the Killing spinors
(\ref{kil}) into eigenspinors of $\g^*$ given in (\ref{eg}) and
(\ref{m2kil2}) one can easily solve (\ref{sucon}).
The situation on 
supersymmetries of the embeddings in (\ref{embed}) may be
summarized as follows:
\bea \label{mtwo+}
M_-2:&& \quad x_0^{A'} =0 \quad  \G_{+-1} \ep_0 = - \ep_0 \quad 16~
\rm{supercharges}\\
M_-2:&&  \quad x_0^{A'} \ne 0 \quad \rm{above~and}~ \G^+\ep_0 =0
\quad 8~ \rm{supercharges} \label{mtwo+x} \\
M_+2:&&        \quad  \qquad    \rm{no~preserved~supercharges}
\label{mtwo-}
\eea

It is guaranteed that supersymmetries of the embedding are
preserved by the kinetic and interaction Lagrangians of
fluctuations (in both $X^m$ and $\q$) about the static
configurations in (\ref{embed}). By construction the embedding
SUSY's satisfy $\delta \q =0$ when $\q=0$. They are therefore
realized linearly on fluctuations. The principal question
investigated below is whether there are new linear fermionic symmetries
of the fluctuations about displaced $M_-2$ branes and $M_+2$
branes which effectively increase the number of supercharges
beyond those counted in (\ref{mtwo+})-(\ref{mtwo-}).

\subsection{Gauge fixing}

The next step of our investigation requires gauge fixing of both
world volume diffeomorphisms and $\kappa$ symmetry. We
encountered some initially puzzling issues of compatibility of
gauge-fixing conditions with the specific embeddings
(\ref{embed}). These issues were not known to the previous
investigators we consulted, so we will describe them in some detail.

It was shown in \cite{DSR, SY} that the plane
wave membrane action is quadratic in fermions in lightcone gauge,
just as it is in flat space \cite{BST}. Here lightcone gauge
consists of the conditions \be \la{lccond} \G^+ \q = 0; \hsp X^+ =
p^+ \t; \hsp g_{\t p} = 0; \hsp g_{\t \t} = - {\rm{det}} g_{pq}
\ee where $(\t, \sigma^p)$ with $p=1,2$ are the worldvolume
coordinates. Note that these conditions do not entirely fix the
worldvolume diffeomorphisms; the group of area preserving
diffeomorphisms remain.

Given the simplicity of the action in lightcone gauge, this gauge
appears at first sight to be the natural choice for us. However,
the embeddings in which we are interested are
degenerate in this gauge. To prove this consider a bosonic embedding
and gauge fix $X^+ = p^+ \t$. Then the induced worldvolume metric is
\bea
\g_{\t \t} &=& - 2 p^+ \pa_{\t} X^- + (\pa_{\t} X^A)^2 + (p^+)^2 G_{++}; \\
\g_{\t p} &=& - p^+ \pa_{p} X^- + \pa_{\t} X^A \pa_{p} X^A; \hsp
\g_{pq} = \pa_{p} X^A \pa_{q} X^A. \nn
\eea
Now impose the next condition from (\ref{lccond}), namely $\g_{\t p} = 0$;
this condition will determine $X^-$ from the remaining scalars
$X^A$ (see \cite{duffetal}) for details).
The final condition in (\ref{lccond}) is needed to remove the 
square root $\sqrt{-\det g_{\m\n}}$ and give a polynomial
action. However, even before imposing this,
one finds 
\be
{\rm{det}} (g_{\mu \nu}) = g_{\t \t} {\rm{det}} (g_{pq}).
\ee
Since the configurations in (\ref{embed}) describe branes
extended in $(X^+,X^-,X^{\tilde{A}})$, there is a single worldvolume direction
$X^{\tilde{A}}$ transverse to the lightcone. 
The induced brane metrics are thus degenerate for our
embeddings, \ie $\det(-g_{\m\n})=0$ and are thus inadmissible
in lightcone gauge.


It may seem surprising that these embeddings are degenerate
in this gauge, given that they are clearly not degenerate in
the static gauge $X^{\pm} = \xi^{\pm}$ and $X^{\tilde{A}} = \xi^{\tilde{A}}$,
which is necessarily related to the lightcone gauge by a worldvolume
diffeomorphism (\ref{wd}). However, the Jacobian of the transformation
from $(\xi^{\pm}, \xi^{\tilde{A}})$ to $(\t, \sigma^p)$ is zero. This follows
from imposing the conditions
\be
\xi^+ = p^+ \t; \hsp
0 = - p^+ \pa_{p} \xi^- + \pa_{\t} \xi^{\tilde{A}} \pa_{p} \xi^{\tilde{A}},
\ee
which enforce $X^+  = p^+ \t$ and $\g_{\t p} = 0$ respectively.

Thus we must give up lightcone gauge for the bosons in favor of
the static gauge which is immediate for our embeddings. However,
we might consider a ``hybrid'' gauge  in which the bosonic static
gauge conditions are combined with  lightcone gauge for the
fermions ($\G^{+} \q = 0$). The matrix  $\cM$  vanishes
in this gauge \cite{DSR} so the supervielbeins  in (\ref{superv})
are quadratic in the fermions. The square root
$\sqrt{-\det g_{\m\n}}$ will still contain terms up to order
$\q^{16}$, but the Lagrangian is still much simpler than for
other fermionic gauges.

However, this hybrid gauge is also singular in the neighborhoods of
our embeddings. To prove that a given fermionic gauge is admissible
one needs to show that there always exists a kappa symmetry
transformation to bring any theta into this gauge. In the case at
hand this requires that
\be
\G^{+} (\q + (1 - \Gamma) \kappa) = 0
\ee
admits solutions for $\kappa$ which remove all 16 (arbitrary)
components of $\q^+$, where $\q^+ = - \half \G^- \G^{+} \q$.

Now let $\Gamma = \gamma^* + \delta \Gamma$ where $\gamma^*$ is
$\Gamma$ evaluated on the classical embedding and $\delta
\Gamma$ contains field fluctuations; then $\gamma^{*2} =1$
and $(1 \pm \gamma^*)$ are projectors of rank 16. For the
embeddings (\ref{embed}),~~
$\gamma^*$ is  $\G_{+-1}$ or $\G_{+-4}$; thus $[\gamma^*, \G^+] =
0$.   These two facts immediately imply
\be
(1 + \gamma^*) \G^{+} (\q - \delta \Gamma \kappa) = 0.
\ee
This means that the kappa transformations needed to remove
the eight components of theta satisfying $\gamma^* \q^{+} = \q^{+}$
are non-perturbative in that $\kappa \sim (\delta \Gamma)^{-1} \q^+$.
The conclusion is that the fermion lightcone gauge is singular
at the embeddings of interest.

The singularity of the hybrid gauge near the embedding is also
manifest on gauge fixing within the functional integral. When one
tries to introduce ghosts for the hybrid gauge fixing one finds
as usual that one needs an infinite number of ghosts for ghosts.
Leaving this well-known problem aside, one also finds that the leading
term in the ghost action is cubic in the fields for 24 out of the 32
ghost components. Thus the ghost action does not admit a traditional
perturbative formulation, which is directly connected to the
observation above that the compensating kappa
transformation is non-perturbative.


Given these problems, we choose to work instead
with the physical gauge, namely $\gamma^* \q = \q$, which can manifestly
always be reached in the neighborhood of the embedding. Given the
complexity of the supervielbeins we will work only to quadratic order
in the fermions. Fortunately this turns out to be sufficient to
investigate the question of enhanced supersymmetry for displaced
$M_-2$ branes and $M_+2$ branes.

\subsection{Action to quadratic order in fermions} \label{2fr}

It is straightforward to compute the supermembrane theory on the plane wave 
supergeometry specified by (\ref{superv}) to quadratic order in the 
fermions. From (\ref{superv})-(\ref{sv})  we obtain
\be
\Pi_\mu^r = \pa_\m X^n e_n^r + \bar{\q} \G^r \td{D}_\mu \q + \co(\q^4)
\ee
where
\be \la{tdd}
\td{D}_\mu \q= \pa_\m \q
+ \pa_\mu X^m (e_m^r T_r + {1 \over 4} \w_m{}^{rs} \G_{rs}) \q
\ee
and $T_r = T_{r}{}^{stuv} F_{stuv}$.
From this expression we obtain $g_{\mu \nu}$ to quadratic order in the fermions
\be \label{m2}
g_{\mu \nu} = \g_{\mu \nu} + 2 \bar{\q} \pa_{(\m} X^n \td{\G}_n \td{D}_{\n)}\q
+ \co(\q^4)
\ee
where $\g_{\mu \nu} = \pa_\m X^m \pa_\n X^n G_{mn}$ is the induced
worldvolume metric and $\td{\G}_n = e_n^r \G_r$ are curved gamma
matrices. Using these results we obtain
\bea \label{act2th}
S &=& -\int d^3 \xi \sqrt{- {\rm{det}} \g_{\mu \nu} }
\left(1 +  \g^{\mu \nu}  \pa_\m X^n \bar{\q} \td{\G}_n \td{D}_{\n}\q
+ \co(\q^4) \right) \\
&&+ \int d^3 \xi \e^{\l \mu \nu} \left(
{1 \over 6} C_{lmn} \pa_\l X^l \pa_\m X^m \pa_\n X^n
- \half  \bar{\q} \td{\G}_{mn} \td{D}_{\l}\q \pa_\m X^m \pa_\n X^n +
\co(\q^4) \right). \nn
\eea
 
\section{$M_{-}2$ branes} \la{m-sec}

We now discuss the case of $M_{-}2$ branes along $(+,-,1)$.  
The physical gauge corresponds to 
\be
X^{\mu} = \xi^{\mu}, \qquad  
\qquad \G_{+-1} \q = \q, 
\ee
where $\mu = (+, -, 1)$.
We are interested in both the brane at the origin and the brane away from the 
origin. Recall that the worldvolume scalars parameterize the 
transverse position of the brane. To obtain the action for the 
brane at the origin we expand around $X^{A'}_{cl} = \theta=0$, 
whereas for the brane localized at $x_0^{A'}$ 
we instead expand around $X^{A'}_{cl} =x_0^{A'} , \theta=0$
where $A'=(i',a)$ runs over all transverse directions.

Clearly the action for the brane at the origin is given by (\ref{act2th})
and to obtain the action for the brane away from the origin we simply 
have to shift $X^{A'}$ by $x_0^{A'}$. It will useful to introduce 
a double grading to count the order in fluctuations and in the constant 
position $x_0$. The action and variation are then split into terms
of definite order and we will denote them as $S^{p}_{q}$ and $\d^p_q$,
where the superscript $p$ denotes the order of fluctuating fields 
and the subscript $q$ denotes the order of the constant positions. 
For instance, the quadratic part 
of the action for the brane at the origin will be denoted by $S^2_0$
etc.

 To explicitly evaluate the action we need to know $\td{D}_\mu \q$
 and $h^{\mu \nu} \equiv \sqrt{- \det \g_{\mu \nu}} \g^{\mu \nu}$:
 \bea 
 &&\td{D}_\mu \q =
 \left(\pa_\mu  + \pa_\mu B 
 + \left({\mu \over 12} (\G_{23} + 2 \G_{123})
 -\qu \pa_A G_{++} \G_{-A}\right)\d_{\m+}
  + {\mu \over 6} \G_{-23} \d_{\mu 1}
 \right) \q;  \nn \\
 && h^{++} = - \td{\g}_{--}; \qquad h^{+1} = \td{\g}_{-1};
 \qquad h^{-1} = \td{\g}_{+1} + G_{++} \td{\g}_{-1};
 \nn \\
 && h^{--} = -G_{++} - \half G_{++} \td{\g}_{11} - \td{\g}_{++}
 -G_{++} \td{\g}_{+-} - \half G_{++}^2
 \td{\g}_{--}; \label{ginv} \\
 && h^{+-} = -1 - \half \td{\g}_{11} - \half G_{++} \td{\g}_{--}; 
 \hsp h^{11} = 1 - \half G_{++} \td{\g}_{--} 
 - \half \td{\g}_{11} - \td{\g}_{+-}, \nn
 \eea
 where  $B={\mu \over 6} (X^2 \G_{-3} - X^3 \G_{-2}- \half X^a \G_{-23a})$, 
 $\td{\g}_{\mu \nu} = \pa_{\mu} X^{A'} \pa_{\nu} X^{A'}$
 and we have only kept terms quadratic in fluctuations.

 \subsection{Quadratic action}

 The action to quadratic order in the fields is
 \bea \la{act}
 S^{2}_{0} = - \int d^3 \xi (1 + \half \g_{(0)}^{\mu \nu} \pa_{\mu} X^{A'}
 \pa_{\nu} X^{A'} + 2 \bar{\q} \td{\G}^{\mu} D_{\mu} \q +
 \mu X^+ (\pa_{+} X^2 \pa_{-} X^3 - \pa_{+} X^3 \pa_{-} X^2) ).
 \eea
 where $\g_{(0)\mu \nu}$ and  $\td{\G}^{\mu}$
 are the fluctuation independent part 
 of the induced metric $\g_{\m \n}$ and 
 $\g^{\m \n} \pa_{\nu} X^n \tilde{\G}_n$, respectively.
 Note that we take $\ep^{+-1} = 1$. Notice that the fermion
kinetic term receives a contribution 
both from the Dirac and the WZ part of the action.
 The Dirac operator appearing in (\ref{act}) can be written as
 \be \la{Dir}
 \td{\G}^{\mu} D_{\mu} = (\G^{-} \pa_{-} + \G^{+} \pa_{+} +
 \G^1 \pa_{1} + \half G_{++} \G^{+} \pa_{-} + \qu \mu \G^{+123} ).
 \ee
 The last term couples the eight physical worldvolume spinors i.e.
 the $SO(8)$ part is not diagonal.

 The action (\ref{act}) describes fluctuations of a brane
 located at the origin in the transverse directions; here 
 $G_{++}=- \frac{\mu^2}{9} (x^1)^2$. To describe fluctuations of
 a brane embedded at constant non-zero transverse position one
 needs to expand instead about $X^{A'} = x_{0}^{A'}$. The resulting
 action to quadratic order in the fluctuations is
 \bea \la{shift}
 S_{{\rm{shifted}}} &=& S^{2}_{0} + S^{2}_{2}; \\
 S^{2}_2 &=& - \half \ca \int d^3 \xi (  (\pa_{-} X^{A'})^2 -
 2 \bar{\q} \G^{+} \pa_{-} \q); \nn \\
 \ca &=&  \frac{\mu^2}{36} \left(4 (x^{i'}_0)^2 + (x^a_0)^2 \right), \nn
 \eea
 where $i' =2,3$.

 Note that (\ref{shift}) can be obtained from (\ref{act})
 provided that in the latter $G_{++}$ is kept exact.
 Although terms $(x^{A'})^2
 (\pa X^{B'})^2$ are clearly subleading (quartic) and do
 not contribute at quadratic order for the brane at the origin
 they do contribute to the quadratic term $(x_0^{A'})^2 (\pa X^{B'})^2$
 appearing in the action for the shifted brane.

 \subsection{Interactions} \label{secint}

 We next compute a subset of the interaction terms.
 For the brane at the origin there are no
 cubic interaction terms; the leading order interactions are quartic.
 The terms quartic in bosonic fluctuations can be obtained straightforwardly
 by expanding the $\sqrt{-\g}$. To obtain the terms quartic in fermion 
 fluctuations one would need to extend the results of section (\ref{2fr}).
 This is a somewhat tedious computation (which could however
 be done using the results of sections \ref{pw} and \ref{m2sec}).
 Fortunately, it is sufficient for our purpose (as we explain below)
 to consider only quartic terms that are quadratic in both bosonic
 and fermionic fluctuations. The Dirac term contributions to these are
 \bea \label{Dirac}
 S^{4}_{0(D)} &=& - \int d^3 \xi \left ( \td{\g}_{--} \bar{\q} \G^- \pa_{+} \q
 - \td{\g}_{+-} \bar{\q} \G^1 \pa_1 \q + (\td{\g}_{+1} + G_{++} \td{\g}_{-1})
 \bar{\q} \G^1 \pa_{-} \q \right .  \\
 && \hsp \hsp  + \td{\g}_{-1} \bar{\q} (-\G^- \pa_1 + \G^1 \pa_{+}) \q 
 + \half \td{\g}_{11} \bar{\q} (\G^- \pa_{-} - \G^1 \pa_1) \q
 + \qu \mu ( \td{\g}_{--} \bar{\q} \G^{-23} \q \nn \\
 &&  \hsp \hsp
 \left .
 + \td{\g}_{-1} \bar{\q} \G^{123} \q) 
 - \half G_{++} \td{\g}_{--} \bar{\q} (\G^1 \pa_1 - \G^- \pa_-) \q
 + \qu \pa_1 G_{++} \td{\g}_{--} \bar{\q} \q  + \cdots \right )  \nn
 \eea
 The relevant Wess-Zumino term contributions are
 \bea \la{wzint}
 S^{4}_{0(WZ)} &=& - \int d^3 \xi \left
 ( \half \e^{\mu \nu \l} \bar{\q} \G^{A'B'}
 \pa_\m \q \pa_{\n} X^{A'} \pa_{\l} X^{B'}
 - {1 \over 12} \mu \pa_{-} X^a \pa_1 X^b \bar{\q} \G^{23ab} \q
 \right . \\
 && \left .
 + \tw \mu [(\pa_1 X^a \pa_{-} X^2 - \pa_- X^a \pa_1 X^2) \bar{\q} \G^{13a} \q
 + 5 (\pa_1 X^2 \pa_- X^3) \bar{\q} \q
 - 2 \leftrightarrow 3 ]
 + \cdots \right ), \nn
 \eea
 where the ellipses in both contributions
 denote terms containing $(\bar{\q} \G^{+} \cdots \q)$
 which are irrelevant in what follows (the ellipses 
 do not contain $\G^-$).

 For the brane at non-zero transverse position there are cubic
 interaction terms. The relevant terms are those which are
 linear in bosonic and quadratic in fermionic fluctuations:
 \bea \label{s31}
 S^3_{1} &=& \int d^3 \xi \left
 (- \cb (  (\pa_- X^{A'})^2 - 2 \bar{\q} \G^+ \pa_- \q)
		  + \half \cb_{B'} (\pa_- X^{A'}) \bar{\q} \G^{+A'B'}
 \q \right ) \\
 \cb &=&  \frac{\mu^2}{36} 
 (4 x_0^{i'} X^{i'} + x_{0}^a X^a); \qquad
 \cb_{B'} =  \frac{\mu^2}{18} (4
 x_{0}^{i'} \delta_{i' B'} + x_{0}^a \delta_{a B'}). \nn
 \eea

 \subsection{Fluctuation spectrum}

 We now use the quadratic actions (\ref{act}) and (\ref{shift})
 to work out the fluctuation
 spectra of the branes. The bosonic equations of motion following from
 (\ref{act}) and (\ref{shift}) are
 \be \label{sceq}
 \Box X^{a} = 0; \hsp
 \Box \phi = - i \mu \pa_- \phi,
 \ee
 where
 \be
 \Box = (- 2 \pa_{+} \pa_{-} - G_{++} \pa_{-}^2 + \pa_{1}^2),
 \ee
 and $\phi = (X^2 + i X^3)$ is a complex scalar. Note that
 the scalars $X^2$ and $X^3$ are coupled by the Wess-Zumino term so one has
 to diagonalise their equations of motion. Here $G_{++} = -(\frac{\mu^2}{9}
 (x^1)^2 +\ca)$ where $\ca$ is given in (\ref{shift}) and
 is zero for a brane at the origin.

The Dirac equation is 
\be \label{dir}
0=\td{\G}^{\mu} D_{\mu} \q =(\G^{-} \pa_{-} + \G^{+} \pa_{+} +
 \G^1 \pa_{1} + \half G_{++} \G^{+} \pa_{-} - \qu \mu \G^{+23} )  \q
\ee
Iterating we get,
\be \label{dsq}
0= \td{\G}^{\nu} D_{\nu} \td{\G}^{\mu} D_{\mu} \q = 
\Box \q + \half (\mu \G^{23} + \pa_1 G_{++} \G^+) \pa_- \q.
\ee
Recall that the 16 $\q$ satisfy $\G_{+-1} \q = \q$.\footnote{
Notice that in later sections we use the notation
$\G_{+-1} \q^\pm = \pm \q^\pm$ and in this notation the 16 $\q$ are 
the $\q^-$ ones. In order to avoid clumsy notation such 
as $\q^{-,\pm}_\pm$ we suppress this superscript below.}
Let us further decompose the fermions into eigenspinors of $\G^1$,
\be 
\G^1 \q^{\pm} = \pm \q^{\pm}.
\ee
Multiplying (\ref{dsq}) by $\G^- \G^+$ yields
\be
\Box \q^- = -\half \mu \G^{23} \pa_- \q^-
\ee
where we used the relations $\G^+ \q^+=0$ and $\G^- \q^-=0$.
Since $\G^1$ commutes with $\G^{23}$ we can further decompose 
$\q^-$ into eigenspinors of $\G^{23}$,
\be
\G^{23} \q^-_{\pm} = \pm i \q^-_{\pm}
\ee
We thus obtain
\be
\Box \q^-_{\pm} = \mp \half i \mu \pa_- \q^-_\pm
\ee
which of the same form as the scalar field equation (\ref{sceq}).
It remains to discuss $\q^+$ components of $\q$. Multiplying 
the fermion field equation (\ref{dir}) by $\G^+$ yields
\be
\pa_- \q^+ = - \half \G^+ \pa_1 \q^-
\ee
Provided $p^+ \neq 0$ this equation determines $\q^+$ from
$\q^-$. Thus there are 8 independent fermion modes. 

 So for both bosonic and fermionic fluctuations we need to solve
 equations of the form
 \be
 \Box \varphi = i c \pa_{-} \varphi,
 \ee
 for various values of $c$.
 Decomposing into Fourier modes along the lightcone, 
 $\varphi = \exp (i p^+ x^- + i p^- x^+) \varphi(x^1)$, this becomes
 \be
 (2 p^+ p^-  -  \frac{1}{9} (p^+)^2 \mu^2 (x^1)^2 + \pa_{1}^2
 - \Delta) \varphi(x^1) = 0,
 \ee
 with
 \be
 \Delta = (p^+)^2 \frac{\mu^2}{36} (4 \sum_{i'} (x_0^{i'})^2 +
  \sum_{a} (x_0^a)^2) - c p^+ \equiv
 2 p^+ \Delta H - c p^+ ,
 \ee
 where the $x_0^{i'}$ and $x_{0}^a$ are the constant transverse
 positions about which the brane is fluctuating.
 Recall that the eigenfunctions of the harmonic oscillator
 satisfy
 \be
 \left(\pa_1^2 +(1+2n- \frac{1}{9} (p^+)^2 \mu^2 (x^1)^2) \right)  H_{n}(x^1)
 =0 
 \ee
The Gaussian part of the Hermite function behaves as
 $\exp (- \frac{1}{6} \mu p^+  (x^1)^2)$
 and decays exponentially. Notice that we take $p^+>0$.  
 Thus the $p^-$ eigenvalue is determined as
 \be \la{lch}
 p^- = (1 + 2n) \frac{\mu}{6} + \Delta H - \frac{1}{2} c.
 \ee

 The spectra of fluctuations are characterized by their $(p^-,n)$
 eigenvalues for a given $p^+$.
 From (\ref{lch}) one finds that the lowest $p^-$
 eigenvalues for given $p^+$ (i.e. those for which $n=0$)
 for the fluctuations are respectively:
 \begin{center}
 $\begin{array} {c c c }
 \phi & & + \half \mu \\
 \q^-_+ & & + \qu \mu \\
 X^{a} & \Delta H + \tq \mu & + 0 \\
 \q^-_- & & - \qu \mu \\
 \bar{\phi} & & - \half \mu \\
 \end{array}$
 \end{center}
 Furthermore raising $n$ by one unit increases the $p^-$ eigenvalue
 by $\mu/3$. This analysis shows that the transverse position enters the
 spectrum only as a universal shift $\Delta H$ in the $p^-$ eigenvalue for
 all bosonic and fermionic fluctuations. Thus the brane
 away from the origin is as supersymmetric as the brane at the origin,
 just as in the corresponding computation of D-brane spectra
 in the maximally supersymmetric IIB plane wave in \cite{ST3}.

\subsection{Worldvolume supersymmetry}

We now discuss in detail the worldvolume supersymmetries of these
branes. The emergence of worldvolume supersymmetry from spacetime
supersymmetry on gauge fixing kappa symmetry was first discussed
in detail in the context of the four-dimensional supermembrane
in \cite{AGIT}. The discussion here follows closely that of \cite{Kal}:
we determine which combined kappa and supersymmetry transformations
leave the gauge fixed action invariant.

Let us split both $\kappa$ and $\ep$ into eigenspinors
of $\G_{+-1}$, defining $\G_{+-1} \lambda^{\pm} = \pm \lambda^{\pm}$
for any spinor $\lambda$. The appropriate splitting of the Killing
spinors is
 \bea \la{eg}
 \ep &=& \ep^- + \ep^+  \\
 \ep^- &=& (1 + \frac{\mu}{6} \G^{+23} x^1) e^{\frac{\mu}{12} x^+ \G^{23}
 - \frac{\mu}{6} x^+ \G^{123} } \ep_{0}^- \nn \\
 && - \frac{\mu}{12} (x^a \G^{a} - 2 x^{i'} \G^{i'}) \G^{+23}
 e^{\frac{\mu}{12} x^+ \G^{23}} \ep_{0}^+ \nn \\
 \ep^+ &= & 
 (1 + \frac{\mu}{6} \G^{+23} x^1) e^{- \frac{\mu}{12} x^+ \G^{23}
 - \frac{\mu}{6} x^+ \G^{123} } \ep_{0}^+ \nn \\
 && + \frac{\mu}{12} (x^a \G^{a} - 2 x^{i'} \G^{i'}) \G^{+23}
 e^{- \frac{\mu}{12} x^+ \G^{23}} \nn
 \ep_{0}^-
 \eea
where  $i'=2,3$.

As discussed, we fix kappa symmetry by setting 
\be
\q^- = 0.
\ee
Notice that the kappa symmetry transformations (\ref{supkap}) 
have their own gauge invariance: taking $\k \to \k + (1+\G)\k'$,
for a local spinor $\k'(x)$, leaves the transformation rules invariant. 
We gauge fix this invariance by setting $\k^+=0$. 

Demanding that a combined kappa symmetry
and supersymmetry transformation preserves the gauge $\q^- = 0$ 
requires that we choose the parameters such that
\be
\kappa^{-} = - \half \ep^{-}.
\ee
The worldvolume supersymmetry transformations are now given by
\bea
\delta \q &=& \e + (1-\G) (- \half \ep^{-})  \\
\d X^{A'} &=& -\bar{\e} \G^{A'} \q - \bar{\q} \G^{A'} (1-\G) 
(- \half \ep^{-}) \nn
\eea
In what follows we will only be concerned with the transformation
rules up to terms linear in the fluctuations. The reason is that 
we are only interested in whether the transformation rules
contain an inhomogeneous term (which means the 
supersymmetry is non-linearly realized) or not.
Notice that the symmetry rules (\ref{supkap}) themselves receive
higher order corrections  in $\q$, as noted at the end of section 3.

The field dependent $\Gamma$ expanded to leading order about the embedding is
 \bea
 \Gamma = \G_{+-1} + \G_{+- A'} \pa_1 X^{A'}
 + \G_{-1 A'} \pa_{+} X^{A'} + \G_{1+ {A'}} \pa_{-} X^{A'}
+ \half G_{++} \G_{-1A'} \pa_- X^{A'}
 \eea
This $\Gamma$ contains all the terms needed to obtain the transformation
rules to linear order in fluctuations. 
The resulting combined transformations are then
 \bea \la{trf}
 \delta \q &=& \half
 \td{\G}^{\mu} \G^{A'} \pa_{\mu} X^{A'} \ep^{-} + \ep^+; \\
 \delta X^{A'} &=& 2 \bar{\q} \G^{A'} \ep^{-}, \nn
 \eea
where $\td{\G}^{\mu}=\g^{\m \n} \pa_\n X^n e_n^r \G_r$ (but only the 
fluctuation independent part contributes) and
$\ep^{+}$  and $\ep^-$ are given in (\ref{eg}). Note that these combined
transformations do not preserve the static gauge (since
$\delta X^{\mu} = \bar{\q} \td{\G}^{\mu} \ep^+$ is non-zero) 
and thus a compensating diffeomorphism (\ref{wd}) with parameter $\eta^{\mu} = - \bar{\q} 
\td{\G}^{\mu} \ep^+$ is needed to maintain the gauge. The latter implies further terms in the combined
transformations (\ref{trf}) which are however always at least quadratic in
the fluctuating fields and can be neglected below. 

 \subsubsection{Symmetries to quadratic order}

 Let us now consider the supersymmetries of the quadratic actions,
 for branes located both at and
 away from the origin. In the former case the action is given in
 (\ref{act}) and in the latter case in (\ref{shift}).

 The explicit form of the symmetries follows from substituting
 the Killing spinors into (\ref{trf}) and keeping terms to
 the appropriate order. This gives
 \bea \label{susy0}
 \d^1_{0} \q &=& \half \td{\G}^{\mu A'} \pa_{\mu} X^{A'}
 (1 + \frac{\mu}{6} \G^{+23} x^1) e^{\frac{\mu}{12} x^+ \G^{23}
 - \frac{\mu}{6} x^{+} \G^{123}} \ep_{0}^- \nn \\
 && + (1 + \frac{\mu}{6} \G^{+23} x^1) e^{-\frac{\mu}{12}
 x^+ \G^{23} - \frac{\mu}{6} x^+ \G^{123} } \ep_{0}^+ \\
 && + \frac{\mu}{12} (X^{a} \G^{a} - 2 X^{i'}
 \G^{i'}) \G^{+23}
 e^{-\frac{\mu}{12} x^+ \G^{23}}
 \ep_{0}^-; \nn \\
 \d^1_0 X^{A'} &=& 2 \bar{\q} \G^{A'} (1 + \frac{\mu}{6} \G^{+23} x^1)
 e^{\frac{\mu}{12} x^+ \G^{23} - \frac{\mu}{6} x^+ \G^{123}} \ep_{0}^-. \nn
 \eea
where the subscript $p$ in the variation $\d_p^q$ denotes the order
of field fluctuations and the superscript $q$ denotes the order of the constant
positions.
Of these transformations the $\ep_{0}^+$ transformations are
clearly inhomogeneous and non-linearly realized (see second line).
The $\ep_{0}^-$ transformations are however linearly realized; these correspond
to the sixteen worldvolume supersymmetries of the brane at the origin
found in the probe analysis. 

Now consider expanding about constant transverse positions;
the action is given in (\ref{shift}).
The supersymmetry transformations follow from shifting the fields
$X^{A'}$ in the previous expressions. To show this explicitly
notice that the action at the origin $S^0$ is of the schematic form
\be \label{s0}
S^0 = \int \sum_n X^n F_n (\q, d X)
\ee
where we suppress spacetime indices and $F_n$ are expressions that 
depend on $\q$ and on derivatives of $X^{A'}$ but not on 
undifferentiated $X^{A'}$. In other words we make explicit in (\ref{s0})
the dependence on undifferentiated $X^{A'}$. Similarly the 
supersymmetry rules are of the form
\be \label{su0}
\delta X = \sum_n X^n G_n (\q,dX,\e), \qquad
\delta \q = \sum_n X^n H_n (\q,dX,\e)
\ee
for appropriate $G_n$ and $H_n$ ($\e$ is the supersymmetry parameter). 
The explicit form of
the lowest order  $F_n, G_n$ and $H_n$ can be read off from (\ref{act}) and
(\ref{susy0}) but we will not need them. We now shift the brane 
away from the origin by setting $X = x_0 + Y$,
where for clarity we call the fluctuating part $Y$,
\be \label{actsh}
S_{{\rm shift}} = \int \sum_n (x_0 + Y)^n F_n (\q, d Y)
\ee
It is simple to show that invariance of (\ref{s0}) under (\ref{su0})
implies that (\ref{actsh}) is invariant under
\be \label{su1}
\delta Y = \sum_n (x_0+Y)^n G_n (\q,dY,\e), \qquad
\delta \q = \sum_n (x_0+Y)^n H_n (\q,dY,\e).
\ee
The issue is whether the new transformation rules contain inhomogeneous
pieces or not. If they do then the corresponding supersymmetries 
will be non-linearly realized.

Let us now return to our specific case.
Notice that to obtain all terms that are at most linear in 
fluctuations for the brane away from the origin we need to know
certain terms that are higher order in fluctuations
for the susy variation for the brane at the origin.
The only quantities that 
contain undifferentiated $X^{A'}$ are $G_{++}$ and spin-connection
$\omega_+{}^{-A'}$, so one has to keep track of the dependence
on them.
 
The transformation rules that depend on $\ep_0^+$ acquire 
new contributions linear in fluctuation but they still contain 
the inhomogeneous term given in (\ref{susy0}). These 
supersymmetries are non-linearly realized and they will not discussed
further. The supersymmetries that depend on $\ep_0^-$ 
are given by 
 \bea
 \delta_{\rm{shifted}} &=& \delta^{1}_{0} + \delta^{0}_{1} + \delta^1_{2}; \\
 \delta^{0}_{1} \q
 &=& \frac{\mu}{12} (x^{a}_0 \G^{a} - 2 x^{i'}_0
 \G^{i'}) \G^{+23}
 e^{-\frac{\mu}{12} x^+ \G^{23}}
 \ep_{0}^-; \label{shsym} \\
 \delta^1_{2} \q &=& -\qu \ca \G^{+A'} \pa_{-} X^{A'}
 e^{-\frac{\mu}{12} x^+ \G^{23}} \ep_{0}^-.
 \eea
where $\ca$ is given in (\ref{shift}). The $\delta^1_{2} \q$ 
term originates from the term in $\d_0^1 \q$ containing
$\tilde{\G}_+=\G_+ - \half G_{++} \G_-$. The supersymmetry transformation 
of $X^{A'}$ remain unchanged. 
The $\ep_{0}^-$ transformations now contain
an inhomogeneous piece for spinors such that $\G^{+} \ep_{0}^- \neq 0$.
Thus only eight worldvolume supersymmetries seem to be linearly realized,
in agreement with the probe analysis.

The analysis so far explicitly confirms general expectations:
the linearly realized supersymmetries are exactly the ones
predicted by (\ref{sucon}). We now show however that for 
the brane away from the origin and to quadratic approximation
in fluctuations the worldvolume theory admits 8 additional
linearly realized supersymmetries. As discussed above,
the invariance of the action at the origin implies a corresponding 
invariant for the brane away from the origin,
\be \label{inv}
(\delta^1_{0} + \delta^0_1 + \delta^1_{2}) (S^2_0 + S^2_2) = 0.
\ee
This leads to a number of relations obtained by collecting terms 
that contain the same number of fields and are of the 
same order in $x_0$:
\bea
&&\delta^1_{0} S^2_{0} = 0; \label{old} \\
&&\delta^0_{1} S^2_0 = 0; \label{news} \\
&& \delta^1_{0} S^2_2 + \delta^1_{2} S^2_0 = 0; \label{hom0} \\
&&\delta^0_{1} S^2_2=0. \label{newse}
\eea
That these relations should hold follows from the general
argument given earlier but we have also explicitly verified 
them. (\ref{old}) is the supersymmetry invariance of the brane 
at the origin. What is important is (\ref{news}) which 
says that the action for the brane at the origin is symmetric
by itself under the inhomogeneous symmetry transformation $\d^0_1$. This 
follows from the fact that the action for the 
brane shifted away from the origin does not contain a term 
linear in $x_0$. Furthermore, this symmetry extends to a symmetry of 
the brane away from the origin, see (\ref{newse}). For this to be true 
it is crucial that the shifted rules do not contain $\d_3$.
The invariance of the shifted action under $\d_0^1$ implies that 
\be \label{hom}
\delta_{\rm{hom}} =\delta^1_{0} + \delta^1_{2}
\ee
is also a symmetry of the shifted action by itself  (as can also 
be verified using (\ref{old}) and (\ref{hom0})). This is 
however a homogeneous transformation generated by all sixteen $\ep_{0}^-$,
so the quadratic approximation of the worldvolume theory for
brane away from origin admits 16 linear supersymmetries.
  
 The invariance of the action under $\d^0_1$ is a special
 case of the ``semi-local'' invariance
 \be \label{sl}
 \delta \q = \G^+ \chi(x^+)
 \ee
 where $\chi(x^+)$ is an arbitrary spinor that depends only
 on $x^+$. One may check that the quadratic part of the action
 is invariant under this transformation. There is also a similar
 bosonic ``semi-local'' invariance
 \be
 \delta X^{A'} = f^{A'} (x^+)
 \ee
 where $f^{A'}(x^+)$ are arbitrary functions of $x^+$.
This additional ``gauge'' invariance allows one to gauge 
away the inhomogeneous term leading to an additional set 
of 8 linear realized supersymmetries.

These considerations explain why the spectrum of the shifted brane is as
 supersymmetric as that of the brane at the origin. Furthermore,
 one can understand the shifted values of $p^-$ for the fluctuations
 of the former as follows. The structure of the superalgebra implies
 that the anticommutation of $\delta_{\rm{hom}}$
 with itself should generate (amongst other terms related to
 the rotation charges) a transformation corresponding to 
 $P^-$.  Since $\{\delta^1_0, \delta^1_2 \} \sim \ca$  and $\{ \delta^1_2,
 \delta^{1}_2 \} = 0$ the $p^-$ values for the brane away from
 the origin are clearly shifted by a term proportional to $\ca$
 as we found.

\subsubsection{Interactions}

We now turn to the question of whether the extra supersymmetries are respected 
by interactions. By construction the full action is invariant
under the symmetries generated by $\ep_{0}^-$.
A sufficient and necessary condition for there to be an extension of the 
homogeneous symmetry (\ref{hom}) to the full theory
is hence that the inhomogeneous symmetry $\d_1^0$ extends to a symmetry of the 
interacting theory. In other words, there should be a 
deformation of $\d_1^0$ (possibly containing fluctuating fields) which
leaves the action of the interacting theory invariant. 

The action for the interacting theory of the shifted brane is
\be
S=S^2_0 + S^2_2 + S^4_0 + S_1^3 + \cdots
\ee 
where (relevant parts of) $S^4_0$ and $S_1^3$ are given in (\ref{Dirac}),
(\ref{wzint}) and (\ref{s31}). We have seen in the previous
section that $\d_1^0 S_0^2=0= \d_1^{0} S^2_2$. Furthermore, it is also true 
by inspection that 
\be
\d_1^0 S_1^3 =0. 
\ee
Let us now discuss $S_0^4$. A direct computation shows that $\d_1^0 S^4_0$
is not zero. In fact the appropriate extension of (\ref{inv}) implies
the relation 
\be \la{hoe}
(\d_1^0 S_0^4 + \d_0^1 S_1^3 + \d_1^{2} S^{2}_0) = 0,
\ee
where the explicit form of $\d_1^2$ (not needed here) follows from 
extending the previous arguments to higher order. One can verify 
explicitly that each term in the above is non-vanishing. 

The issue is however whether there is an appropriate deformation
of $\d_1^0$,
\be
\d = \d_1^0 + \sum_{p,q} \tilde{\d}_p^q,
\ee
for appropriate $p$ and $q$, that leaves the interacting action invariant. 
Notice that $\d_1^0 S^4_0$ contains
3 fluctuating fields and is linear in the constant positions.
This means that terms of the same order as $\d_1^0 S^4_0$
are only produced by $\tilde{\d}_0^1 S_1^3$ and $\tilde{\d}_1^2 S_0^2$
(just as in (\ref{hoe})). In order not to upset the lowest order 
invariance (i.e. the invariance of $S_0^2$) the new transformation should be 
at least quadratic in fluctuations. The only possible deformation
$\td{\d}_0^1$ which leaves invariant the lowest order action
is $\d_0^1$ itself, but such a deformation leads us back to
the original supersymmetry variations. 

We are thus led to look for a variation
$\tilde{\d}_1^2$ such that 
\be
\d_1^0 S^4_0 + \tilde{\d}_1^2 S_0^2 =0, \quad \Leftrightarrow \quad
\d_1^0 S^4_0 = - {\delta S_0^2 \over \delta Z^M} \tilde{\d}_1^2 Z^M
 \quad \Leftrightarrow \quad
\d_1^0 S^4_0 \approx 0
\ee
where $\approx$ means equality when the lowest order equations hold.
We thus obtain that a necessary and sufficient condition for
$\d_1^0$ to be extendible to a symmetry of the leading interactions is that 
$\d_1^0 S^4_0$ vanishes weakly.

Notice that $\d_1^0$ does not mix terms with a different number of 
fermions. This means that the variation of terms in $S_0^4$ that are quadratic
in $\q$ should vanish separately from the variation of terms quartic
in $\q$. Clearly the terms containing $(\bar{\q} \G^{+} \dots \q)$,
where the ellipses do not contain $\G^-$, are trivially 
invariant under $\d_1^0$ (because $(\G^+)^2=0$). Thus we only 
need to examine the remaining terms; this is the reason 
why only these terms were listed in section \ref{secint}. Now notice that 
the structure of the bosonic fluctuation terms in the Dirac and 
Wess-Zumino part of $S_0^4$, equations (\ref{Dirac}) and 
(\ref{wzint}) respectively, is different: the latter is 
antisymmetric under the exchange of two bosons whilst the former
is symmetric. This means that the two sets of terms cannot 
mix with each other under the $\d_1^0$ variation, except possibly 
through the use of lowest order field equations. The latter, however,
are diagonal for $X^a$ ($a=4,...,9$). It follows that a necessary 
condition for the extension of $\d_1^0$ (and thus
of the extra supersymmetries) into a symmetry of the 
leading interactions is that the $\d_1^0$ variation of the terms 
in (\ref{Dirac}) that do not depend on $X^2$ and $X^3$ vanishes weakly.

Explicit computation yields
 \bea \label{remain}
 \d_1^0 S_D &=& - \int d^3 \xi \left (2 \bar{\chi'} \G^+
 (\pa^2_- X^{a} \G^- + \pa_1 \pa_- X^{a} \G^1) \q X^{a} \right. \\
 && \left. \hsp \hsp \hsp
 + \half \pa_- (X^{a} X^{a}) \bar{\chi'} \G^+ (\G^- \pa_- + \G^1 \pa_1) \q
 - \pa_- X^{a} \Box X^{a} \bar{\chi} \G^+ (\G^- + \G^1) \q \right) \nonumber
 \eea
where $\chi$ is defined by $\d_1^0 \q = \G^+ \chi$,  $\d_1^0 \q$
is given in (\ref{shsym}), and $\chi' = (\pa_+ + {\mu \over 4} \G^{23}) \chi$.
The last line is proportional to the lowest order field equations.
To check whether the term in the first line is a total derivative 
up to lowest order field equations we use the lemma that a term 
is a total derivative of a local field 
if and only if its Euler-Lagrange derivative\footnote{
The Euler-Lagrange derivative of a local function $f$ is defined as 
$\delta f/\delta \f = \pa f/\pa \f 
- \pa_\mu (\pa f/\pa (\pa_\mu \f)) + \cdots$.} 
with respect to all fields vanishes (for a proof see, for instance, 
section 4.4 of \cite{physrep}). One finds by inspection that the 
Euler-Lagrange derivative 
with respect to $\q$ is non-vanishing (even when the free field 
equations hold). We therefore conclude that the variation
$\d_1^0 S_0^4$ does not vanish weakly and thus that the extra supersymmetries 
are not respected by interactions.

 \section{$M_{+}2$ branes} \la{m+sec}

 We consider in this section an $M_{+}2$-brane oriented along 
$(+,-,4)$ and compute the
 action to quadratic order in fermions in the physical gauge
 $\G_{+-4} \q = \q$. One can use the general expression given in
 section 2, along with explicit expressions
 \be
 \td{D}_{\mu} \q = \left( \pa_{\mu} + \pa_{\mu} B  
 + \left(\frac{\mu}{12} (\G^{-+123} + 2 \G^{123})
 - \qu \pa_{A} G_{++} \G_{-A} \right) \d_{\mu +} 
 - \frac{\mu}{12} \G_{4} \G^{+123} \d_{\mu 4}
 \right)\q.
 \ee
 where $a' = 5,..,9$ and $B=\frac{\mu}{6} (X^{i} \G_{i} \G^{+123}
 - \half X^{a'} \G^{+123a'})$. Substituting these expressions into the action
 one finds the following action to quadratic order
 \be
 S^{2} = - \int d^3 \xi (1 + \half \g_{(0)}^{\mu \nu} \pa_{\mu} X^{A'}
 \pa_{\nu} X^{A'} + 2 \bar{\q} \td{\G}^{\mu} \pa_{\mu} \q),
 \ee
 where $A'$ labels the eight transverse scalars and $\g_{(0)\mu\nu}$
 and $\td{\G}^{\mu}$ are as given for the $M_{-}2$-brane. Note
 that there is no coupling of the fluctuations to the background
 RR flux to quadratic order. This leads to a degeneracy in
 the spectrum; the equations of motion are
 \be
 \Box X^{A'} = 0; \hsp
 \td{\G}^{\mu} \pa_{\mu} \q = 0,
 \ee
 and thus following the previous analysis one finds that all
 eight physical bosons and fermions have a ground state
 energy of $\Delta H + \tq \mu$. As previously advertised this
 spectrum is akin to that of the $D_{+}$ branes
 considered in \cite{ST2,ST3}.

 The target space supersymmetries are realised to this
 order on the brane as
 \be
 \delta \q = \half \td{\G}^{\mu} \G^{A'} \pa_{\mu} X^{A'} \ep^{-}
	      + \ep^{+}; \hsp
 \delta X^{A'} = 2 \bar{\q} \G^{A'} \ep^{-},
 \ee
 where the relevant splitting is now $\G_{+-4} \ep^{\pm} =
 \pm \ep^{\pm}$. Decomposing the Killing spinors one finds
 \bea
 \ep^+ &=& \left( \cos (\frac{\mu}{6} x^+) \cos (\frac{\mu}{12} x^+)
-\G^4 \sin (\frac{\mu}{6} x^+) \sin (\frac{\mu}{12} x^+) \right. \nn \\
&& \left. - \frac{\mu}{12} \G^+ 
[(x^{a'} \G^{a'} - 2 x^{i} \G^{i})\G^{123} \cos (\frac{\mu}{12} x^+)
+ x^4 \sin (\frac{\mu}{12} x^+)] \right) \ep_{0}^+ \label{m2kil1} \\
&+&\left( \G^{123} [-\sin (\frac{\mu}{6} x^+) \cos (\frac{\mu}{12} x^+)
+\G^4 \cos (\frac{\mu}{6} x^+) \sin (\frac{\mu}{12} x^+) \right. \nn \\
&& \left. - \frac{\mu}{12} \G^+ 
[(x^{a'} \G^{a'} - 2 x^{i} \G^{i})  \sin (\frac{\mu}{12} x^+)
- x^4 \G^{123} \cos (\frac{\mu}{12} x^+)] \right) \ep_{0}^- \nn
\eea
 and 
 \bea
 \ep^- &=& \left(-\G^{123} [\sin (\frac{\mu}{6} x^+) \cos (\frac{\mu}{12} x^+)
+\G^4 \cos (\frac{\mu}{6} x^+) \sin (\frac{\mu}{12} x^+) \right. \nn \\
&& \left. + \frac{\mu}{12} \G^+ 
[(x^{a'} \G^{a'} - 2 x^{i} \G^{i})  \sin (\frac{\mu}{12} x^+)
- x^4 \G^{123} \cos (\frac{\mu}{12} x^+)] \right) \ep_{0}^+ \label{m2kil2} \\
 &+& \left( \cos (\frac{\mu}{6} x^+) \cos (\frac{\mu}{12} x^+)
+\G^4 \sin (\frac{\mu}{6} x^+) \sin (\frac{\mu}{12} x^+) \right. \nn \\
&& \left. - \frac{\mu}{12} \G^+ 
[(x^{a'} \G^{a'} - 2 x^{i} \G^{i})\G^{123} \cos (\frac{\mu}{12} x^+)
+ x^4 \sin (\frac{\mu}{12} x^+)] \right) \ep_{0}^- \nn
\eea
The splitting indicates that all 32 target space
supersymmetries are realised
non-linearly regardless of the brane location; this follows from
the inhomogeneous first terms in the first and third lines of $\ep^+$
and $\ep^-$.

Unlike the case studied in the previous section, we do 
not find in this case extra symmetries that can be used 
to remove the inhomogeneous parts of the transformations.
Furthermore, the spectrum suggests that the supersymmetries of
the $M_+2$-branes should not be directly related to the Killing
spinors. This follows from the fact that the symmetries of
the spectrum manifestly commute with the lightcone Hamiltonian, according
to the degeneracy found above, whilst the target space supercharges
do not commute with the lightcone Hamiltonian \cite{Chr,Fig}. Also in 
the corresponding string analysis in \cite{ST2,ST3}
the restored kinematical symmetries were not directly
related to the target space kinematical symmetries.

These considerations lead us to the following homogeneous
symmetries of the action
\be
\delta^{0} \q = \half \td{\G}^{\mu} \G^{A'} \pa_{\mu} X^{A'} \ep_{0}^-; \hsp
\delta^{0} X^{A'} = 2 \bar{\q} \G^{A'} \ep_{0}^-, \la{sym}
\ee
where the constant parameter $\ep_{0}^-$ satisfies
$\G^{+} \ep_{0}^- = 0$. This gives eight linearly realised
symmetries irrespective of the brane location. Note also
that these symmetries are directly analogous to the restored
kinematical symmetries of the $D_{+}$ branes discussed in \cite{ST2, ST3}.

To determine whether interactions respect these symmetries one
needs to compute the action to next order. In this case there
are cubic interaction terms given by
\bea
S^{3} &=& \half \mu \int d^3 \xi (\pa_{-} X^{a'} \bar{\q} \G^{a'123} \q
- \pa_{-} X^{i} \bar{\q} \G_{i} \G^{4123} \q -
\pa_{4} X^{i} \bar{\q} \G_{i} \G^{+123} \q ) \\
&& - \mu \int d^3 \xi x^+ \ep^{\mu \nu \rho} (\pa_{\mu} X^1
\pa_{\nu} X^2 \pa_{\rho} X^3). \nn
\eea
Now suppose that the symmetry (\ref{sym}) can be extended to
the full action. This implies that there is a transformation
$\delta = \delta^{0} + \delta^{1} + ...$ such that
\be
\delta^{0} S^{3} + \delta^{1} S^{2} = 0,
\ee
which can only be satisfied for some $\delta^{1}$ (to be determined)
provided that $\delta^{0} S^{3} \approx 0$ up to terms which
vanish on-shell with respect to the lowest order field equations.

However, $\delta^{0} S^{3}$ contains the term
\be
\delta^{0} S^{3} 
= - \half \mu \int d^3 \xi (\bar{\q} \G^{4 a' b' 123} \ep^{-}_0
(\pa_{4} X^{b'} \pa_{-} X^{a'}) + \cdots ).
\ee
This is the only term in the variation which contains both $X^{a'}$
and $X^{b'}$ (with $b' \neq a'$) and thus cannot cancel against
other terms. It cannot however be written either as a total
derivative or as a term proportional to the lowest order field
equations, which can be most easily seen 
using the Euler-Lagrange test of the previous section, and 
thus it violates the symmetry.

\section{Discussion} \la{dis}

We have studied in this paper the worldvolume supersymmetries for
M2 branes in the maximally supersymmetric plane wave 
background of M-theory. 
%
By construction, the M2 worldvolume theory 
has as many supersymmetries as the background. 
However, only a subset of them are linearly realized. 
We explicitly constructed all worldvolume supersymmetries
using standard methods and found results in agreement with 
the probe analysis, i.e. the number of linearly 
realized supersymmetries matched the probe results.

We showed, however, that the quadratic approximation 
to the worldvolume theory admits additional 
linearly realized supersymmetries. In the case 
of $M_-2$ branes localized away from the origin this came 
about by the use of an extra ``semi-local'' invariance
that the quadratic action possesses. This ``gauge''
invariance allows one to gauge away the inhomogeneous 
term from certain supersymmetry rules, thus providing 
8 additional linearly realized supersymmetries.
In the case of $M_+2$ branes the additional supersymmetries
were completely new symmetries, unrelated to target 
space supersymmetries. In both cases, we explicitly computed the 
spectrum of small fluctuations and showed that it is
organized into multiplets of the additional supersymmetries.
In both cases we showed that the new supersymmetries 
are not respected by the worldvolume interactions.

This investigation was motivated by the analysis in 
\cite{ST1,ST2,ST3} of D-branes on the maximally supersymmetric
plane wave background of IIB string theory. The probe analysis \cite{ST1}
leads to results analogous to those for M2 branes. 
Moreover as in the cases studied here 
the open string spectrum appears to be organized by 
more supercharges than given by the probe analysis. 
Corresponding additional exceptional supercharges
were constructed in \cite{ST2,ST3} as additional Noether 
charges. The additional fermionic symmetries 
depend crucially on special properties of the plane wave background 
and the fact that the string worldsheet was considered to be a strip. 

The results in \cite{ST1,ST2,ST3} imply that the D-brane
worldvolume theory should admit more supersymmetries,
albeit only at the quadratic level if the interactions 
break the extra symmetries. In this paper we found that 
the M2 branes do exhibit extra supersymmetries but 
only at the quadratic order. The mechanism of supersymmetry
enhancement mimics the corresponding string theory 
construction. Our results strongly suggest that the 
extra symmetries in the case of D-branes are not respected by 
string interactions and one could verify this by repeating 
the computation described here for the worldvolume theory of 
D-branes.

An interesting question is whether the branes studied here 
as well as the corresponding D-branes in the IIB plane
wave are stable. Consider the case of 
$M/D_-$ branes localized away from the origin.  
These branes have the same mass density as 
the corresponding branes at the origin, 
i.e. the worldvolume $\sqrt{-\det \g}$ is independent 
of the position of the brane, and there is no classical 
force that acts on them. However, their excitations 
have an excess of energy that depends on the position 
of the brane. This suggests that brane will emit 
the extra energy and recoil to the origin. In the 
case of D-branes a possible decay channel is via
emission of closed strings\footnote{We thank J. Maldacena
for discussions about this point.}. Since the position 
of the D-brane is arbitrary one may consider the brane
localized very far from the origin so that the 
excess energy is very large and semi-classical methods
may be applicable. 


It would be interesting to understand under which conditions
the quadratic part of the worldvolume theory admits more 
supersymmetries and the spectrum is more supersymmetric
than one would naively expect. It seems likely that there
is corresponding supersymmetry enhancement in other pp-wave
backgrounds, given the similar generic form of the Killing 
spinors. 

Since interactions do not
respect the extra supersymmetries, the masses of the states
are expected to be split by quantum corrections. As masses
are mapped to conformal dimension in the gravity/gauge theory
correspondence this would translate into a prediction for 
anomalous dimensions.

\section*{Acknowledgments}
We would like to thank E.~Bergshoeff, E.~Sezgin, W. Taylor and 
J-T.~Yee for discussions. The research of DZF is supported by the 
National Science Foundation
Grant PHY-00-96515 and by the DOE under cooperative research agreement
DF-FC02-94ER40818. KS is supported by NWO.

\appendix

\section{Conventions}

The gamma matrices satisfy $\{ \G^r, \G^{s} \} = 2 \eta_{rs}$ where
$\eta_{rs} $ is the tangent space metric. They can be taken to be real in
the Majorana representation. Gamma matrices with multiple indices
denote antisymmetrized products with unit strength. The
Dirac conjugate is defined by $\bar{\psi} = i \psi^{t} \G^0$ for
a generic spinor $\psi$. Here $\G^{0}$ is the charge conjugation matrix
which satisfies $(\G^0)^t = - \G^0$. An important property of gamma matrices
in $D= 11$ is that the matrix $\G^{0} \G_{\a_1 .. \a_p}$ is symmetric
for $p=1,2,5$ and antisymmetric for $p=0,3,4$ (the cases $p > 5$ are
related by duality to these).

A useful explicit basis for gamma matrices
in terms of $SO(2,1) \times SO(9)$ matrices is:
\be
\G^{\pm} = \g^{\pm} \otimes \gamma, \hsp
\G^{1} = \g^{1} \otimes \gamma, \hsp
\G^{A'} = 1_2 \otimes \gamma^{A'},
\ee
with $\g^1 = \sigma^3$, $\sqrt{2} \g^{\pm} = (i \sigma^2 \pm \sigma^1)$ and
\be
\gamma^{A'} = \pmatrix { 0 & \sigma^{A'} \cr
(\sigma^{A'})^{t} & 0 }, \hsp
\gamma = \pmatrix {1_{8} & 0 \cr 0 & - 1_{8} }.
\ee
Here the $\gamma^{A'}$ form a real representation of $SO(8)$.
With these choices $\G_{+-1} = 1_2 \otimes \gamma$. Note
that $\sqrt{2} \G^{\pm} = (\G^{0} \pm \G^{10})$.

\end{document}